\documentstyle[sprocl,epsf,fleqn]{article}
\def\hbar{{\mathchar'26\mkern-7muh}}

\renewcommand{\vec}[1]{\mbox{\boldmath$#1$}}
\newcommand{\Dp}[2]{\frac{\partial #1}{\partial #2}}

\newcounter{dafigcounter}
\def\thedafigcounter{\arabic{dafigcounter}}
\newdimen \figleftportion
\newdimen \figrightportion

\newcommand{\partfig}[4]{\refstepcounter{dafigcounter}
\figrightportion=\textwidth
\figleftportion=#3
\advance \figleftportion by \epsfxsize
\advance \figrightportion by -\figleftportion
\advance \figrightportion by -4mm
\noindent
\begin{minipage}[b]{#3}
\mbox{\epsffile{#1.eps}}
\end{minipage}\hfill
\begin{minipage}[b]{\figrightportion}
\label{#1}\footnotesize Figure \thedafigcounter :\ #2
\vspace*{#4}
\end{minipage}}

\newcommand{\widefig}[2]{\refstepcounter{dafigcounter}
\noindent
\begin{minipage}{\textwidth}   
\begin{center}
\mbox{\epsffile{#1.eps}}
\end{center}
\label{#1}\footnotesize Figure \thedafigcounter :\ #2
\end{minipage}\vskip 5mm}

\newcounter{databcounter}
\def\thedatabcounter{\arabic{databcounter}}
\newcommand{\widetab}[3]{\refstepcounter{databcounter}
\noindent
\begin{minipage}{\textwidth}   
\begin{center}
#1
\end{center}
\label{#2}\footnotesize Table \thedatabcounter :\ #3
\end{minipage}\vskip 5mm}

\def\be{\begin{equation}}
\def\ee{\end{equation}}
\def\bea{\begin{eqnarray}}
\def\eea{\end{eqnarray}}

\begin{document}
\title{COLLECTIVE FLOW IN HEAVY ION REACTIONS AND 
       THE PROPERTIES OF EXCITED NUCLEAR MATTER}
\author{%
J.\ KONOPKA, S.A.\ BASS, M.\ BLEICHER, M.\ BRANDSTETTER,\\ 
C.\ ERNST, L.\ GERLAND, W.\ GREINER, S.\ SOFF, C.\ SPIELES,\\
H.\ ST\"OCKER \footnote{invited speaker at 1$^{\rm st}$ 
Catania Relativistic Ion Studies 1996},
H.\ WEBER, L.A.\ WINCKELMANN}
\address{Institut f\"ur Theoretische Physik\\
         Johann Wolfgang Goethe-Universit\"at\\
         Postfach 11 19 32\\
         D--60054 Frankfurt am Main, Germany\\
         e-mail: konopka@th.physik.uni-frankfurt.de}
\maketitle\abstracts{Quantum Molecular Dynamics (QMD) calculations 
of central collisions between heavy nuclei are used to study fragment
production and the creation of collective flow. It is shown that the 
final phase space distributions are compatible with the expectations from
a thermally equilibrated source, which in addition exhibits a collective 
transverse expansion. However, the microscopic analyses of the transient
states in the intermediate reaction stages show that the event shapes are more
complex and that equilibrium is reached only in very special cases but 
not in event samples which cover a wide range of impact parameters as it is
the case in experiments. The basic features of a new molecular dynamics
model (UQMD) for heavy ion collisions from the Fermi energy regime up to 
the highest presently available energies are outlined.}

\section{Introduction}
The only possibility, how excited nuclear matter can be probed in the
laboratory are nucleus--nucleus reactions~\cite{Sto86}.
In particular when two heavy ions 
like Au or Pb collide most centrally, the combined system forms a zone of
high density and high random agitation of the involved constituents. 
However, it is still an open question, to which extent the system equilibrates
and hence allows for the application of thermodynamical 
concepts~\cite{Kuhn93,Kon94,Stachel}.

From the experimental point of view it is clear that the measurable final
state has to be compatible with an equilibrium configuration. Two basically
different approaches are commonly used: i) analysis of the final spectra
in terms of emission from a thermally equilibrated source, i.e.\ they should
fall off as $\propto \exp(-E/T)$ and ii) analysis of the final fragment 
composition in terms of a chemically equilibrated source, i.e.\ the 
population of a state $j$ is $\propto \exp(\mu_j/T)$, where the 
chemical potentials of all states are connected via some 
Gibbs equilibrium conditions.

For the theory there are many other ``observables'', which may indicate 
the degree of equilibration reached. In this article we use the 
Quantum Molecular Dynamics model (QMD) to analyze the complete space time 
history of heavy ion collisions on a microscopic basis. The conclusions, which
are drawn from the final state of the reaction only are confronted with 
that information, which is obtained from the knowledge about the intermediate
reaction stages. 

\section{Quantum Molecular Dynamics}
Quantum Molecular Dynamics~\cite{Aic88,Pei89,Faes93}
is a dynamical model which calculates the
time evolution of a heavy ion collision in the entire many-body 
phase space. Nucleons are represented as gaussian wavefunctions
\be
\label{QMDWF}
\varphi_j(\vec{x}_j) = \left(\frac{2\alpha}{\pi}\right)^{\frac{3}{4}}
\exp\left\{-\alpha \Big(\vec{x}_j - \vec{r}_j(t)\Big)^2 +
\frac{{\rm i}}{\hbar} \vec{p}_j(t) \vec{x}_j \right\}
\ee
and their centroids are propagated according to the canonical equations
of motion
\be
\label{EminBed}
\dot{\vec{r}_j} = \Dp{H}{\vec{p}_j}  
\quad {\rm and} \quad
\dot{\vec{p}_j} = -\Dp{H}{\vec{r}_j}\,,
\qquad j=1,\ldots, A \,.
\ee
If two particles approach too close in configuration space, so that
\be d_{jk} \le \sqrt{\sigma/\pi} \ee
a hard collision is carried out in a similar fashion as in intranuclear
cascades.
The final state of each binary collision is checked to 
obey the Pauli-principle, otherwise the collision will not be performed
-- in other terms, the collision is Pauli-blocked. 

\section{Collective flow effects in heavy ion collisions}
Collective flow effects in heavy ion collisions are known as a 
sensitive probe of compression of nuclear matter. For semicentral
collisions the bounce-off
in the reaction plane and the squeeze-out
of matter perpendicular to the reaction plane have been predicted
by nuclear hydrodynamics~\cite{Sto86} 
and were later on observed at the Bevalac~\cite{Dos86,Gut89}. 

Meanwhile there have been numerous experimental as well as
theoretical studies on impact parameter-, mass-, and bombarding 
energy-dependence of these phenomena. 
Presently there are two systematic studies of sideward flow
excitation functions under way: The Ni+Ni system will be studied 
jointly by the INDRA and FOPI collaboration from
30 MeV/nucleon to 2 GeV/nucleon (GANIL--SIS energies).
The Au+Au system is presently under
intense investigation in the 1--4 GeV/nucleon energy regime with the
EOS TPC at the AGS~\cite{Part95}. 
The event shapes are very complex and only
triple differential cross sections are suited for a complete 
characterization of such collisions.

However, the strongest
compression effects and therefore also the most prominent flow effects 
are expected in collisions at vanishing 
impact parameter~\cite{Kon94,Dan95,Roy}
Since all particles are participating in the reaction, also the possibility
of reaching thermalization is highest in this case.
By definition, in these collisions the directed component
of sidewards flow vanishes. The expansion is azimuthally symmetric
around the beam axis and may be divided into two components: a symmetric, but
collective expansion of the excited nuclear matter towards transverse
direction and an additional transverse momentum component due to the non-zero
temperature. This is illustrated in Fig.\ \ref{crisgrep}.
The directed transverse
momentum is close to 0 for grazing reactions, and rises continuously until it
reaches its maximum at 4--5 fm for Au+Au, whereafter it drops to 0 at b=0.   
On the contrary, the average transverse momentum is monotonously related to 
the impact parameter and can therefore be used as an impact parameter measure.

In the following we concentrate on the extreme case of b=0, where all
nucleons involved undergo a rapid sequence of hard collisions which 
is a necessary condition for the approach of thermal equilibrium starting
from a nonequilibrium configuration. 

Fig.\ \ref{wilder_pt} 
shows transverse momentum spectra of various charged fragments
obtained with QMD for the system Au (150 MeV/nucleon, b=0) + Au
(symbols) together with fits to these calculated data, which are based on the
assumption of a thermalized source 
of constant $T(r)$ and $\varrho(r)$ up to a maximum $r=R$
which exhibits an additional azimuthally
symmetric transverse expansion with a linearly increasing expansion (flow)
velocity profile, $v_{\perp}(r_{\rm t}) = a\cdot r_{\rm t}$, see~\cite{Kon94}.
In fact, the corresponding count rates have been 
fitted, rather than the invariant distributions, which are displayed.
All spectra seem to be compatible with ``temperatures''
between 20 and 25 MeV and averaged collective flow velocities of 0.10--0.13 c. 
Similar temperatures have been obtained from fits to experimental spectra
of the EOS~\cite{Lisa94} and FOPI-collaboration~\cite{Cof94}. 
The large multiplicities
of heavy fragments measured in the very same systems, however, suggested much 
lower temperatures on the order of 8 MeV~\cite{Kuhn93}. 
This long-standing puzzle will be resolved below.

The collectivity is more emphasized in the spectra 
of the heavy clusters, since the thermal energy 
per nucleon drops as $1/A$. This is 
expressed in Fig.\ \ref{wilder_e}, where the averaged kinetic energy 
in the center of mass frame and the thermal energy per nucleon is plotted
as a function of the fragment mass. The thermal energies obtained from the
fit to the spectra show the expected fall off, whereas the total energy 
is shifted roughly by a constant amount
due to the collective expansion. Note: for light ejectiles like
protons the collective energy amounts to only 30\% of the total energy,
for the heavy fragments the collective--thermal energy sharing is
reversed (70--80\% of the kinetic energy is collective).

Another interesting aspect of the many-body dynamics
can be read off Fig.\ \ref{wilder_e}:
The average available kinetic energy per nucleon in the center of mass
after the reaction, which is indicated by the horizontal line,
is considerably lower than the averaged kinetic energies of free nucleons.
This excess energy can be associated with the heat released due to the 
formation of heavy fragments, which move with much lower kinetic energy
per nucleon as compared to the average. The description of this
behaviour is out of scope of one-body models like BUU/VUU, since they
lack the many-body correlations, which are responsible for the formation of
complex fragments. Thus these models also underpredict the averaged proton
kinetic energies at low bombarding energies.

In order to study a macroscopic piece of nuclear matter, one can define
a central reaction volume: here we take all cells where the local 
density exceeds half the maximum density at this instant. 
In Fig.\ \ref{wilder_times} the properties of the excited
matter inside this volume are displayed as a function of time.
Even in the late stages of the expansion this zone still contains 
$\approx 1/3$ of the mass of the entire Au+Au system. The evolution of
the widths of the local velocity distributions suggests a rapid cooling,
which goes in line with a fast density decrease. The different
temperatures tend to converge in the late stages only, therefore perfect
thermodynamic equilibrium is not reached and the temperatures are much lower
than the flow fit to the spectra would suggest.

At densities around 0.1--0.5 $\varrho_0$, where the freeze-out of fragments
is expected to happen, the corresponding temperatures have dropped below
10 MeV. This is in agreement with the ``chemical temperatures'' employed
for understanding the large intermediate mass fragment multiplicities.

Now the question arises what is wrong with the assumptions underlying  
the global fit to the spectra in Fig.\ \ref{wilder_pt}.
Assuming an overall thermally equilibrated
source at some temperature $T$
which disintegrates not only due to the thermal pressure but also due to
some additional collective expansion, the probability
of finding a particle which has some collective velocity
$v_{\rm coll.}$, i.e.\ ${\rm d}N/{\rm d}v_{\rm coll.}$ is essentially
unknown. It can be expressed as
\begin{equation}
\frac{{\rm d}N}{{\rm d}v_{\rm coll.}} = \frac{{\rm d}N}{{\rm d}r}
\cdot \frac{{\rm d}r}{{\rm d}v_{\rm coll.}}
= \frac{{\rm d}N}{{\rm d}r} \cdot
\left(\frac{{\rm d}v_{\rm coll.}}{{\rm d}r}\right)^{-1}\, .
\end{equation}
Fig.\ \ref{wilder_assump} shows
the shape of the local flow velocity profile.
It rises linearly with $r_{\rm t}$, 
in accord with prescription used for the spectral fit. 
The temperature $T(r_{\rm t})$ (not shown) is in fact reasonably constant
over the central volume.
However, Fig.\ \ref{wilder_assump} shows that the 
density is not at all constant but rather resembles a Gaussian profile!
This trivial statement has drastic consequences: The high transverse
momentum components of the particle spectra do not correspond to the  
high momentum tails of a hot source, but to fastly moving 
cooled cells~\cite{Kon94,Matt95}
The microscopic analysis suggests that the combinations of
density and temperatures, which are traversed in the course of the expansion,
are in agreement with the expectations from expansion along curves of
constant entropy per baryon as used in 
the quantum statistical analysis
of the fragment distributions in the final state \cite{Kuhn93,Hahn88}.

\section{Extension of QMD towards relativistic and ultrarelativistic energies}
The Quantum Molecular Dynamics approach has been recently
generalized for simulations of the dynamics
of heavy ion reactions over a broad range of energies, from 20 MeV/nucleon
up to the highest presently available energies around 
200 GeV/nucleon~\cite{uqmdcode}.
A large variety of baryonic and mesonic states and string excitations
have been incorporated to cover continuously varying physics phenomena
occurring over these four orders of magnitude in energy.
The baryons, mesons, and resonances which can be populated in UQMD are
listed in Tab.\ \ref{tabelle}. Note that all charge conjugate states 
(antiparticles) are included and are treated on the very same footing
as the particles. For higher mass excitations a string picture is employed. 
All states listed below can be produced in string decays, or in s-channel
collisions, and decays of resonances.

The UQMD code allows for systematic studies of excitation functions over a wide
range of energies in a unique way: the basic concepts and the physics
input used in the calculation are the same for all energies. 
A relativistic cascade is applicable over the entire range of energies.
A preliminary molecular dynamics scheme using a hard Skyrme interaction
is used between 100 MeV/nucleon and 4 GeV/nucleon.\\[1mm] 

As an example excitation functions and scaling studies of baryonic
stopping and transverse flow have been performed within the framework of the
new UQMD model.

Baryonic stopping is  
a necessary condition for the creation of dense and highly excited nuclear
matter~\cite{Ren84}. The key observable
is the rapidity distribution of baryons which 
is displayed in Fig.\ \ref{stopp} for three presently used
heavy ion accelerators. In all cases a system as heavy as Au+Au or Pb+Pb
exhibits a gaussian rapidity distribution peaking at midrapidity. However,
the physical processes associated are different: A shift in one unit
of rapidity corresponds to completely different amounts of energy. Thus 
the average longitudinal
momentum loss in the SIS energy regime is mainly due to the creation of 
transverse momentum whereas at the AGS/SPS energy abundant particle production
eats up a considerable amount of the incident energy. 

The creation of transverse flow is strongly correlated with the 
underlying equation of state~\cite{Sto86}. 
In particular it is believed that 
secondary minima as well as the quark-hadron phase transition lead to a 
weakening of the collective sideward flow. The occurrence of the
phase transition should therefore be observable through abnormal
behaviour (e.g.\ jumps) of the strength of collective motion of the 
matter~\cite{Rischke}. 
Of course, UQMD in its present form does not include
any phase transition explicitly. For the purely hadronic scenario 
the averaged in plane transverse momenta for Ni+Ni and Au+Au
in the 0.1--4 GeV/nucleon region are displayed in Fig.\ \ref{pxdirau-ex}. 
Calculations employing a hard equation of state (full symbols) are compared to 
cascade simulations (open symbols). In the latter case only a slight mass 
dependence is observed. For the calculation with potentials the integrated 
directed transverse momentum push per nucleon is more than twice as high for
the heavier system which corroborates the importance of a non-trivial
equation of state of hadronic matter.

The amount of directed transverse momentum scales in the very same way as 
the total transverse momentum produced in the course of the reaction. The
$p_{\rm x}(y)$ therefore depend only on the reaction geometry but not on the
incident energy. This is demonstrated through Fig.\ \ref{au-ypxscal},
where the mean $p_{\rm x}$ as a function of the rapidity
divided by the average transverse momentum of all particles is plotted.

\section{Summary}
Collective motion of excited nuclear matter resulting in very complex
event shapes is a well established phenomenon in heavy ion reactions. 
Triple differential cross sections
only provide sufficient information to characterize the events. 
Even in head on collisions particle emission is non-isotropic and
strong collective effects are present. Moreover, thermal equilibrium is 
established in very special cases and locally only. Slope parameters obtained
from fits to the final momentum spectra give a wrong impression about the 
temperatures even if collective effects have been taken into account.

\section*{Acknowledgments}
This work has been supported by the Bundesministerium f\"ur 
Bildung und Forschung (BMBF), the Deutsche Forschungsgemeinschaft (DFG), and
the Gesellschaft f\"ur Schwerionenforschung mbH (GSI).

\newpage
\widetab{
\mbox{
\setlength{\unitlength}{1cm}
\begin{picture}(9,5.1)
\put(0,2.48){{\small
\begin{tabular}{cccccc}
 \hline  \hline \\[-3mm]
{\normalsize N}&{\normalsize  $\Delta$}%
&{\normalsize $\Lambda$}&{\normalsize $\Sigma$}%
&{\normalsize $\Xi$}&{\normalsize  $\Omega$}\\  \hline
 ${938}$&$ {1232}$&  ${1116}$& ${1192}$& ${1317}$& ${1672}$\\
 ${1440}$& ${1600}$& ${1405}$& ${1385}$& ${1530}$&\\
 ${1520}$& ${1620}$& ${1520}$& ${1660}$& ${1690}$&\\
 ${1535}$& ${1700}$& ${1600}$& ${1670}$& ${1820}$&\\
 ${1650}$& ${1900}$& ${1670}$& ${1790}$& ${1950}$&\\
 ${1675}$& ${1905}$& ${1690}$& ${1775}$&$$&\\
 ${1680}$& ${1910}$& ${1800}$& ${1915}$&$$&\\
 ${1700}$& ${1920}$& ${1810}$& ${1940}$&$$&\\
 ${1710}$& ${1930}$& ${1820}$& ${2030}$&&\\
 ${1720}$& ${1950}$& ${1830}$&$$&&\\
 ${1990}$&         & ${2100}$&$$&&\\
&& ${2110}$&$$&&\\
\hline\hline
\end{tabular}}}
\put(4.5,0.93){{\small
\begin{tabular}{ccccc}
\hline \hline \\[-3mm]
$0^-$ & $1^-$ &$ 0^+$ &$ 1^+$ &$ 2^+$ \\ \hline 
 $\pi$ & $ \rho$ & $ a_0$ & $ a_1$ & $ a_2$ \\
 $K  $ &$   K^*$ & $ K_0^*$ & $ K_1^*$ & $ K_2^*$\\
 $\eta$&$  \omega$& $ f_0 $&  $f_1$ & $ f_2 $\\
 $\eta'$&  $\phi $&  $\sigma$ & $ f_1'$& $ f_2'$\\ 
\hline\hline
\end{tabular}}}
\end{picture}}
}{tabelle}{List of baryons, mesons, and resonances which are included in
the UQMD model. In addition all charge conjugate states are treated on the
same footing. For higher mass excitations meson- and (anti)baryon-strings
are included.}

\newpage
\epsfxsize = 4.55cm %
\widefig{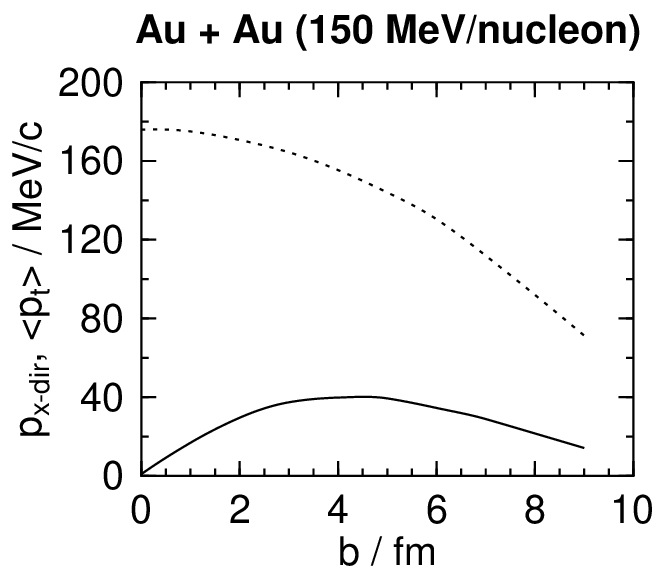}
{Averaged total and in-plane transverse momentum as a function  
of the impact parameter in Au (150 MeV/nucleon) + Au reactions.}

\epsfxsize = 5.81cm 
\widefig{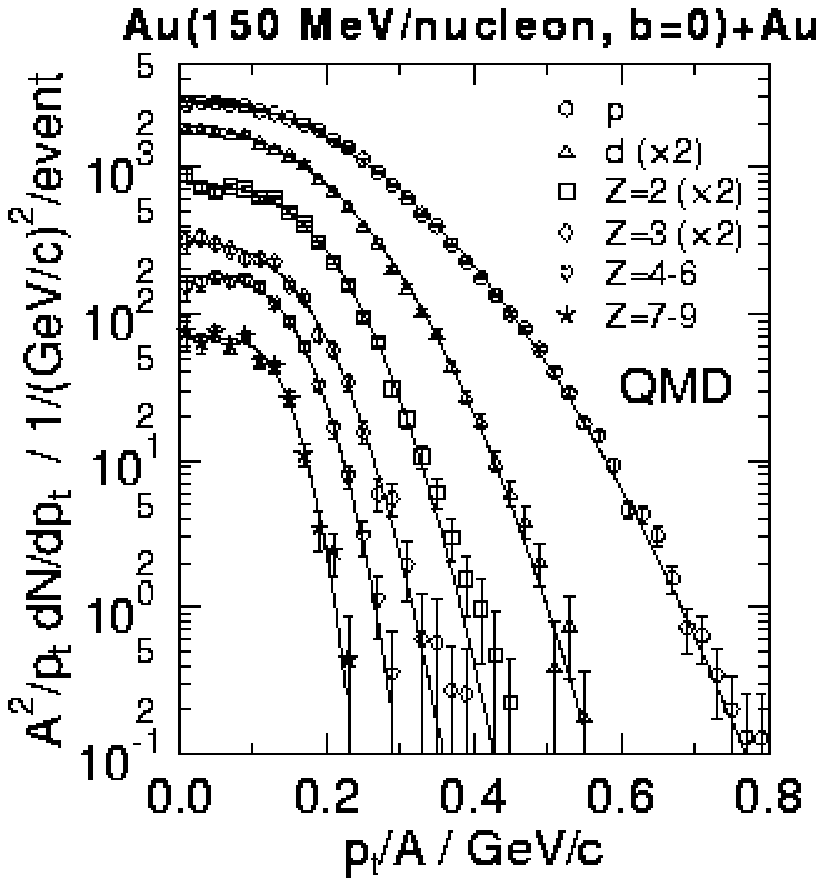}
{Invariant transverse velocity spectra of various 
reaction products of Au (150 MeV/nucleon) + Au at vanishing impact parameter.
The predictions of the Quantum Molecular Dynamics model (symbols) have
been fitted with a thermally equilibrated source, which expands azimuthally
symmetric towards the transverse direction (lines). Note, all spectra are
compatible with temperatures between 20 and 25 MeV and an averaged transverse
collective velocity of 0.10--0.13 c.}

\epsfxsize = 4.06cm 
\widefig{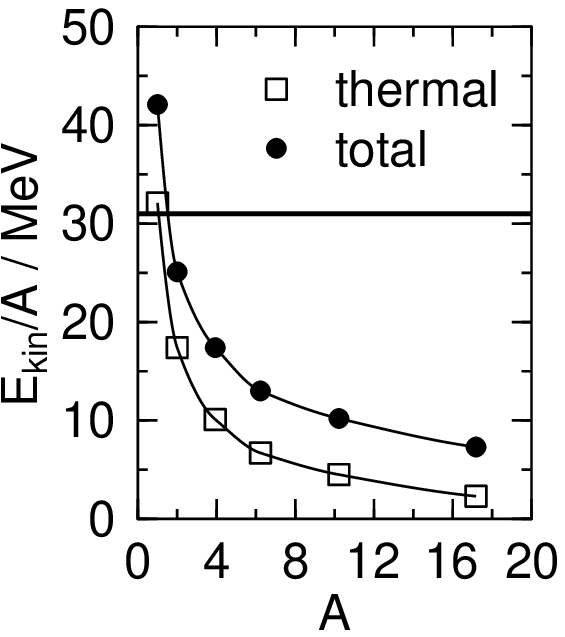}{Averaged kinetic energies in the center of
mass frame of central Au+Au collisions at 150 MeV/nucleon bombarding energy.
The total energy (closed circles) clearly proves a non-thermal behaviour.
The horizontal line indicates the kinetic energy per nucleon in the final 
state averaged over all fragments.}

\epsfxsize = 4.34cm 
\widefig{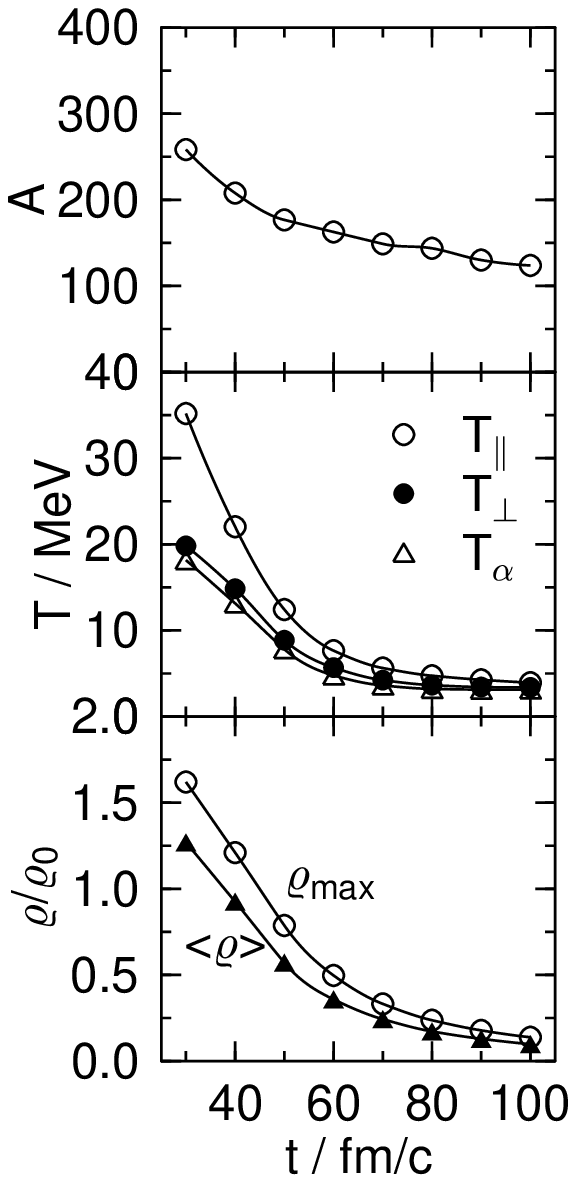}
{Thermodynamics in the central reaction zone, i.e.\ the volume
where the local density is at least half of the maximum value. Mass
content a), widths of the local velocity distributions b), and maximum
as well as averaged density c) are displayed as a function of time.}

\epsfxsize = 9.1cm 
\widefig{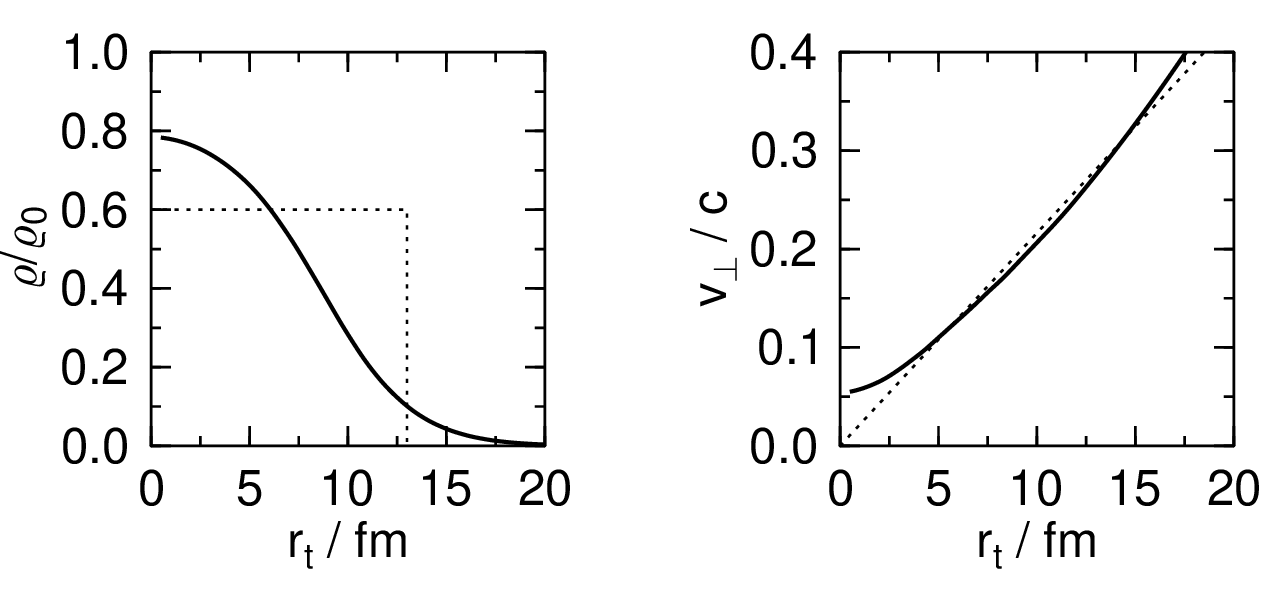}
{Local density and collective velocities as a function of the
transverse distance in the $z=0$ plane. The Figure corresponds again
to the situation at 50 fm/c. The calculated shape (solid lines) are
compared to the assumptions which entered the global fit (dotted lines).
The longitudinal as well as the azimuthal collective velocity vanish.
$v_\perp$ does not reach 0 for $r_t=0$ due to the fact
that the innermost cell is not symmetric around the beam axis.}

\epsfxsize = 5.6cm 
\widefig{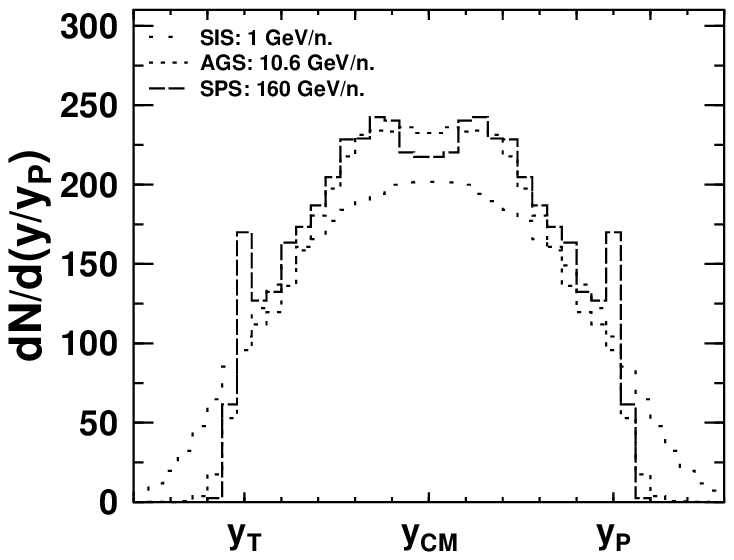}
{Rapidity distributions for Au+Au collisions at SIS (1 GeV/nucleon),
AGS (10.6 GeV/nucleon) and CERN SPS energies (160 GeV/nucleon). All 
distributions have been normalized to the projectile rapidity in the 
center of mass frame.}

\epsfxsize = 5.6cm 
\widefig{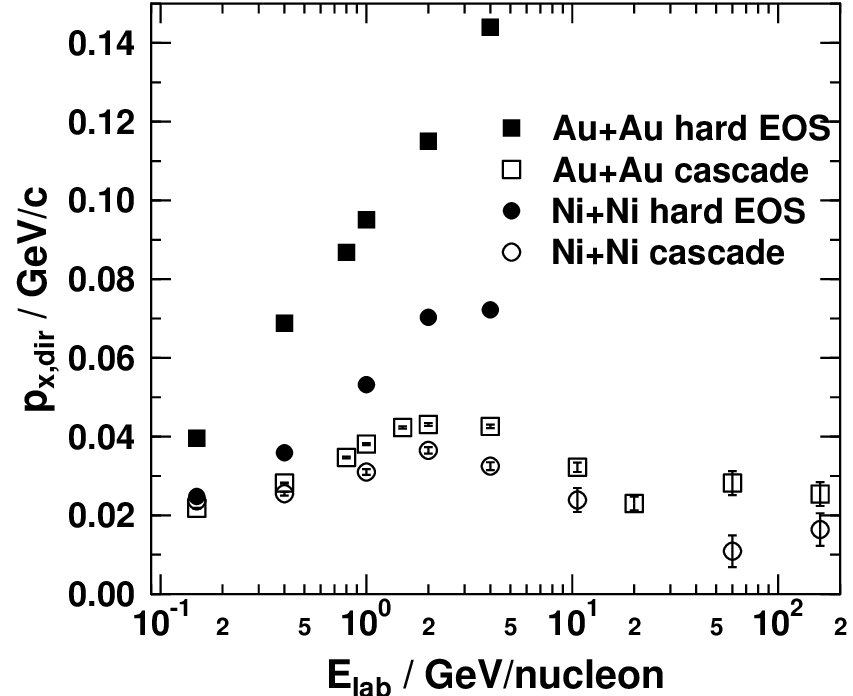}
{Excitation function of the total directed transverse momentum transfer
(px-dir) for the Au+Au and Ni+Ni systems. UQMD results
including a hard equation of
state (full symbols) are compared to the predictions of cascade calculations
(open symbols).}

\epsfxsize = 5.6cm 
\widefig{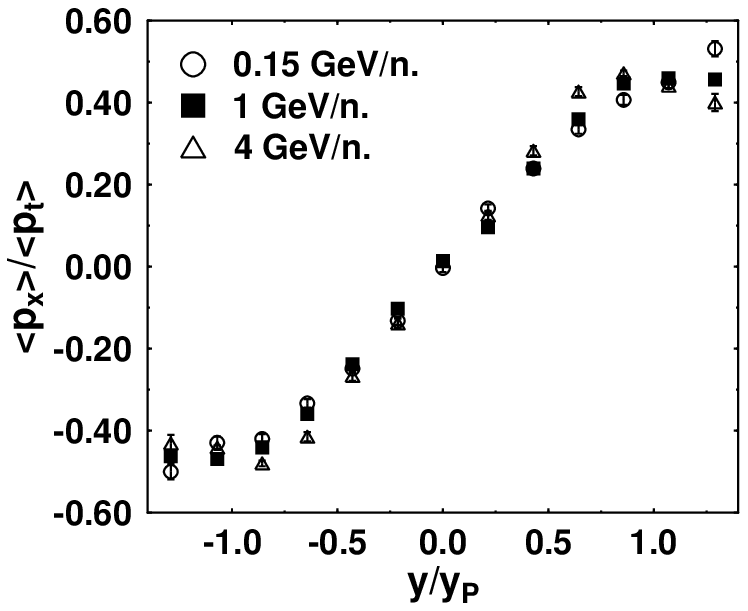}
{Mean directed transverse momentum as a function of the scaled rapidity.
The calculations are performed for Au+Au at a fixed impact parameter of 4 fm.
If the transverse flow is scaled with the mean transverse momentum,
the directed flow does not depend on the bombarding energy.}

\end{document}